\documentclass{PoS}
\def\bml{\begin{mathletters}}
\def\eml{\end{mathletters}}
\def\beq{\begin{equation}}
\def\eeq{\end{equation}}
\def\bea{\begin{eqnarray}}
\def\eea{\end{eqnarray}}

\def\e{{\rm e}}
\def\tr{{\rm tr}}

\def\r{{{\rm r}}}
\def\ii{{{{\rm i}}}}
\def\i{{\hat{{\rm i}}}}
\def\x{{\hat{{\rm x}}}}
\def\y{{\hat{{\rm y}}}}
\def\z{{\hat{{\rm z}}}}

\def\to{\rightarrow}
\def\<{\left\langle}
\def\>{\right\rangle}

\title{
\begin{picture}(0,0)(0,0)%
  \put(350,60){\makebox(0,0)[l]{\textnormal
{\normalsize SU-HET-05-2013}}}%
 \end{picture}%
Critical statistics at the mobility edge of QCD Dirac spectra
}

\ShortTitle{Critical statistics at the mobility edge of QCD Dirac spectra}

\author{\speaker{Shinsuke M. Nishigaki}\thanks{
SMN is supported by Japan Society for the Promotion of Sciences (JSPS)
KAKENHI Grant Number 25400259. 
}\\
       Graduate School of Science and Engineering\\
       Shimane University, Matsue 690-8504, Japan\\
       E-mail: \email{mochizuki@riko.shimane-u.ac.jp}}

\author{Matteo Giordano, Tam\'as G. Kov\'acs, Ferenc Pittler\thanks{MG, TGK
    and FP are supported by the Hungarian Academy of Sciences under
    ``Lend\"ulet'' grant No. LP2011-011.
    TKG and FP acknowledge partial support
    by the EU Grant (FP7/2007 -2013)/ERC No. 208740.}\\ Institute for Nuclear
  Research of the Hungarian Academy of Sciences\\ Bem t\'er 18/c, H-4026
  Debrecen, Hungary\\ E-mail: \email{giordano@atomki.mta.hu,
    kgt@atomki.mta.hu, pittler@atomki.mta.hu}}

\abstract{
We examine statistical fluctuation of eigenvalues from the
near-edge bulk of QCD Dirac spectra above the critical temperature.
For completeness we start by reviewing on the spectral property of Anderson
tight-binding Hamiltonians as described by nonlinear $\sigma$ models and
random matrices, and on the scale-invariant intermediate spectral statistics
at the mobility edge.  By fitting the level spacing distributions,
deformed random matrix ensembles
which model multifractality of the wave functions typical of the Anderson localization
transition,
are shown to provide an excellent effective
description for such a critical statistics.

Next we carry over the above strategy for the Anderson Hamiltonians to the
Dirac spectra.  For the staggered Dirac operators of QCD with 2+1 flavors of
dynamical quarks at the physical point and of SU(2) quenched gauge theory, 
we identify the precise location of the mobility edge as the scale-invariant
fixed point of the level spacing distribution. 
The eigenvalues around the mobility edge are shown to obey critical statistics
described by the
aforementioned deformed random matrix ensembles of unitary and symplectic classes.
The best-fitting deformation parameter for QCD at the physical point turns out to be
consistent with the Anderson Hamiltonian in the unitary class.

Finally, we propose a method of locating the mobility edge at the origin of
QCD Dirac spectrum around the critical temperature,
by the use of individual eigenvalue distributions of deformed chiral random matrices.
}

\FullConference{31st International Symposium on Lattice Field Theory - LATTICE 2013\\
		July 29 - August 3, 2013\\
		Mainz, Germany}

\begin{document}

\section{Introduction}
In this article we shall discuss an intricate parallelism between
the two lattice models
introduced independently by two Nobel laureates, P.W. Anderson and the late K.G. Wilson,
in order to account for physical systems that are apparently unrelated, at least superficially:
doped insulator and strong interaction.
The bridge between the two proves to be a universal framework for quantum energy levels
proposed by another laureate E.P. Wigner, called random matrix theory (RMT) \cite{Meh}.

Historically, it is the statistical fluctuation of lattice Dirac eigenvalues at the `hard edge'
of the spectrum, i.e., near the origin, that has been most extensively investigated,
because it allows one to directly access the effective chiral Lagrangian that governs the
low-energy, nonperturbative regime of the theory, with a help of chiral RMT \cite{SV}.
For one instance, measurement of the pion decay constant via
the statistical response of small Dirac eigenvalues
under the imaginary chemical potential \cite{DHSS,Nis12} provides a
powerful alternative to the conventional method using
the temporal correlation of axial-vector current and pion operators.
On the other hand, the local fluctuation of QCD Dirac eigenvalues
in the {\it spectral bulk} has attracted lesser attention,
as it is not directly connected to the chiral Lagrangian. 
Halasz and Verbaarschot \cite{HV} nevertheless went on to investigate the bulk spectral correlations 
of the staggered Dirac spectrum from dynamical simulation. They concluded that
both short- and long-range correlations of Dirac eigenvalues in the scaling regime
(and less surprisingly in the strong coupling regime)
are perfectly described by the classical RMT of Wigner-Dyson classes.
It indicates that entire Dirac eigenstates are extended throughout the whole lattice sites
just as in the Anderson Hamiltonian with sub-critical randomness.
This finding seems to have lead them to presume
a possibility of Anderson localization in QCD at or above $T_c$, as spelled out in
the very final sentences of their paper which read (quote):
\begin{quote}
"A final point of interest
$\cdots$
is the fate of level correlations during the chiral phase transition.
From solid state physics we know a delocalization transition is associated with a transition in the level statistics
which raises the hope that such phenomena can be seen in QCD as well."
\end{quote}
This statement sounded rather daring,
because of an obvious and essential difference between 
the Anderson tight-binding Hamitonian and the lattice Dirac operator
(beyond whether the disorder sits on or off the diagonal of the matrix):
randomness in Anderson Hamiltonian are mutually independent,
whereas stochastic gauge field variables in QCD are strongly correlated with their neighbors.
To phrase the issue more specifically:
although in their `ordinary phases',
the spectral statistics of both systems allow for a description
in terms of nonlinear $\sigma$ models (NL$\sigma$Ms)
\cite{Weg,Efe} that reflect the global symmetries of operators in concern,
and the above mentioned difference between the microscopic theories could be irrelevant,
the effect of QCD temperature on the spectral $\sigma$ model is unpredictable
as it is not restricted by any symmetry argument.
The purpose of this article, nevertheless, is to draw a definite and affirmative conclusion to the above statement,
from large-scale lattice simulations.
We hope our conclusions of identifying the QCD phase transition as
Anderson localization transition at the zero virtuality, 
and of establishing the presence of localized states in the high-temperature phase,
provide a novel viewpoint of the issue,
especially in the advent of quark-gluon plasma formation by the heavy ion collision.\\

In Sect.\ 2 
we start from basic fact on 
the level statistics of Anderson Hamiltonians 
and the RMT.
We overview the critical statistics at the mobility edge in Sect.3 and
its effective description in terms of deformed RM ensembles in Sect.\ 4.
In Sect.\ 5 we examine the large ensembles of
staggered Dirac spectra of high-temperature QCD with 2+1 quarks at the physical point,
on lattice sizes up to $48^3\times 4$.
Through fitting the local level statistics to the deformed RMs,
we confirm the existence of the scale-invariant mobility edge in the Dirac spectra.
In Sect.\ 6 we propose a strategy of locating the mobility edge at the origin of
Dirac spectrum by the use of individual eigenvalues.

\section{Anderson Hamiltonians and random matrices}
Anderson tight-binding Hamiltonian on a $d$-dimensional lattice $L^d$
with or without a magnetic field is defined as
\cite{And,HS93,HS94},
\beq
H=\sum_\r \varepsilon_\r  a_\r^\dagger a_\r
+\sum_{\langle \r,\r' \rangle} V_{\r\r'}a_\r^\dagger a_{\r'},\ \ \ 
V_{\r,\r\pm\x}=\e^{\mp i \alpha {\rm r}_y},\ 
V_{\r,\r\pm\y}=V_{\r,\r\pm\z}=1, \label{AH12}
\eeq
and with spin-orbit coupling as \cite{EK},
\beq
H=\sum_{\r}\sum_{s=\pm} \varepsilon_{\r}  a_{\r s}^\dagger a_{\r s}
+\sum_{\langle \r,\r' \rangle} \sum_{s,s'=\pm} 
V_{\r s,\r' s'}a_{\r s}^\dagger a_{\r' s'},\ \ \ 
V_{\r s,\r\pm\i\,s'}=
\left(\e^{\mp i\theta {\sigma}_{\ii}}\right)_{s s'}.
\label{AH4}
\eeq
Here $\varepsilon_\r$ are i.i.d.\ random variables on the lattice sites $\r$,
modelling the impurities in the crystal,
and the constants $\alpha, \theta \in [0, 2\pi)$ in the hopping terms
parameterize the strength of the external magnetic field and
the spin-orbit coupling. 
The Hamiltonian matrices (\ref{AH12}) at $\alpha=0$ and (\ref{AH4}) at $\theta>0$ satisfy
real-symmetric and quaternion-selfdual conditions, respectively, whereas (\ref{AH12}) at $\alpha>0$ satisfies
no such (pseudo)reality condition and thus is merely complex-Hermitian.
These three cases are said to belong to
the orthogonal, symplectic, and unitary classes,
and are assigned the Dyson indices $\beta=1,4,2$.
These one-particle Hamiltonians are tailored to model the properties of energy eigenvalues and
eigenstates of electrons in a disordered metal.
For the site energies $\varepsilon_\r$ taken from the Gaussian distribution,
one can analytically perform the ensemble averaging of the $n$-fold product of characteristic polynomials
$Z(\{\lambda\}):=\<\prod_{k=1}^n\det (\lambda_k-H)\>$.
After introducing the Hubbard-Stratonovich field and taking the thermodynamic limit to 
integrate out
its heavy components,
one can derive
a NL$\sigma$M of the universal form \cite{Weg},
\beq
Z(\{\lambda\})=\int_{{\cal M}}DQ\,\exp\left\{
\frac{\pi}{L^d\Delta} \int d^d{\rm r}\, \left( \frac{D}{4} \tr\left| \nabla Q (\r)\right|^2 + i\,\tr\,\Lambda Q(\r)\right)
\right\}.
\label{Z}
\eeq
Here $Q(\r)$ is the soft component of the Hubbard-Stratonovich field
representing the embedding 
of the Nambu-Goldstone manifold ${\cal M}$ associated with the symmetry class
(quaternionic-, complex-, and real-Grassmannian manifold
for $\beta=1,2,4$) in $U(n)$,
$\Lambda={\rm diag}\{\lambda_k\}_{k=1}^n$,
$D$ the diffusion constant that depends on the randomness, and
$\Delta$ the mean level spacing of an eigenvalue window in the spectrum
from which the cluster of $\lambda$s are taken.
Correlation function of densities of states, $\rho(\lambda)={\rm tr}\,\delta(\lambda-H)$,
follows from $Z(\{\lambda\})$ through the replica trick
${\rm tr}\,\delta(\lambda-H)=\lim\limits_{n\to 0}1/(n\pi) \Im{m}\, \partial_\lambda \det(\lambda-i0-H)^n$.
If we choose this window away from the band edges
such that the eigenvalues are populated densely enough and
the typical difference of eigenvalues are much smaller than the Thouless energy defined as $E_c= D/L^2$, 
the path integral is dominated by its zero mode,
\beq
Z(\{\lambda\})\sim\int_{{\cal M}}dQ\,\exp\left( i\,\pi \,\tr\,\frac{\Lambda }{\Delta} Q\right).
\label{Z0}
\eeq
This zero-dimensional NL$\sigma$M would also have followed had we started out from the ensemble of
even-simpler dense matrices $H=(H_{ij})$ distributed according to the Gaussian measure $dH\,\exp(-\tr\, H^2)$
\cite{VWZ}.
For these RM ensembles, known as GOE, GUE, GSE for $\beta=1,2,4$, 
the distribution of spacings of adjacent levels normalized (unfolded)
by the mean level spacing,
$s:=(\lambda_{i+1}-\lambda_{i})/\Delta$,
are analytically calculable in limit of large matrix size \cite{JMMS,TW94}.
For unitary ensembles it is generally expressed 
in terms
of Fredholm determinant, $P_{\beta=2}(s)=\partial_s^2 {\rm Det}(I-K\chi_{[0,s]})$
of the integration kernel $K(x,x')$ (the local asymptotic form of $Z(\lambda,\lambda')$ in (\ref{Z0})),
over the Hilbert space of $L^2$ functions on an interval $[0,s]$.
Those for orthogonal and symplectic ensembles are similarly expressed in terms of 
$2\times 2$ matrix valued kernels,
and are known to be related to the corresponding $P_{\beta=2}(s)$ \cite{TW96,DF}.
In the case of bulk correlation of GUE,
the pertinent kernel is the sine kernel,
\beq
K_0(x,x')=\frac{\sin \pi(x-x')}{\pi(x-x')},
\label{sin}
\eeq
which comprises of trigonometric functions originated from (\ref{Z0}).
The asymptotics of $P_{\beta}(s)$ analytically obtained as the above
takes the `Wigner surmised' form
\beq
P_{\beta}(s)\sim {\rm cst.}\,
s^\beta\ \ (s\ll 1),\ \ \ \log P_\beta(s)\sim -{\rm cst.}\, s^2\ \ (s\gg 1).
\label{Ps-asympt}
\eeq
The level repulsion (for small $s$) and the rigidity (for large $s$) in (\ref{Ps-asympt}) are 
consequences of extended eigenfunctions that are highly likely for a randomly-generated dense matrix.
Thus the numerically-observed perfect agreement between the level spacing distributions (LSDs) $P(s)$ of
Anderson Hamiltonians and random matrices at relatively small randomness and in the spectral bulk
(exemplified in Fig.1) is well understood, 
signifying the extendedness of the single-electron wave functions in a weakly disordered metal.
\begin{figure}[h]
\centering
\includegraphics[scale=0.82,bb=0 0 260 170]{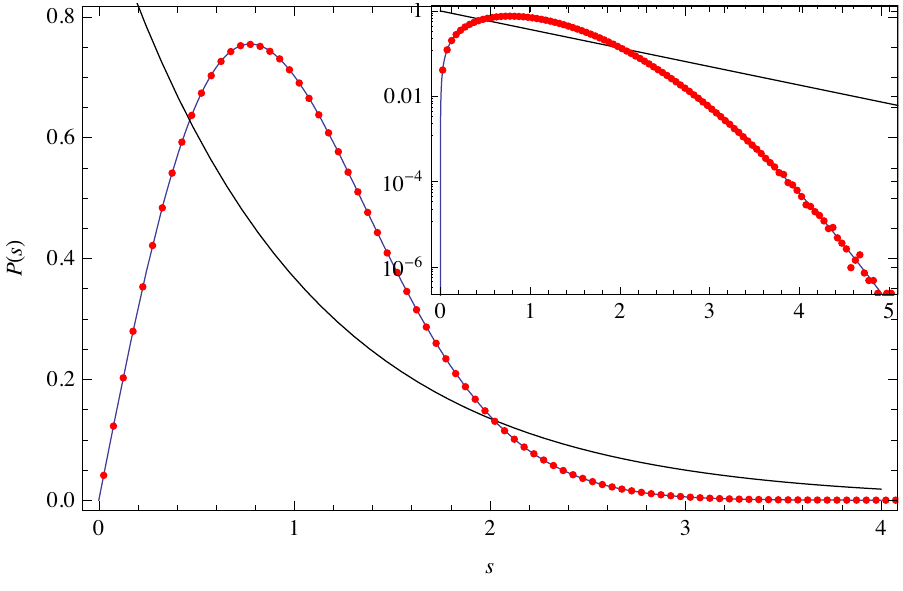}~~
\includegraphics[scale=0.82,bb=0 0 260 170]{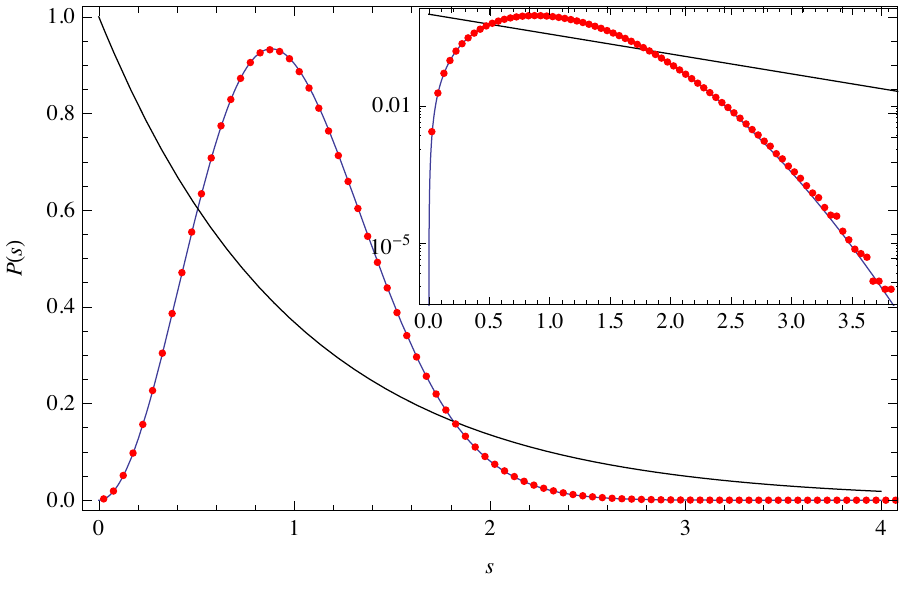}\\
\caption{
Linear and logarithmic (inset) plots of distributions $P(s)$ of unfolded
level spacings of Anderson Hamiltonians at their band center and with weak randomness,
of orthogonal (left) and unitary (right) classes (red dots),
and of GOE and GUE (blue curves).
Poisson distribution ${\rm e}^{-s}$ is plotted in black.
Parameters of Anderson Hamiltonians are: 
lattice size $L^3=20^3$, \#{}samples $=2\cdot10^4$,
with uniform randomness of width $W=4$, $\lambda\in[-2,2]$ (orthogonal), and
$W=10$, $a=1.25$, $\lambda\in[-2,2]$ (unitary).}
\end{figure}

\section{Critical statistics at the mobility edge}

The reduction to the zero-dimensional NL$\sigma$M (\ref{Z0}), i.e.,
to the universality class of RMT,
might not take place for various reasons:
if one simply set the eigenvalue window too close to the band edge,
the typical difference of $\lambda$s could be of the same order or larger than the Thouless energy.
Alternatively, increasing the disorder strength would diminish the diffusion constant, and
accordingly the Thouless energy below the mean level spacing of the energy window in concern.
In either case the condition $\Delta\ll E_c$ is violated and one has to deal with the
whole path integral nonperturbatively, which appeared to be an unfeasible task at the time of perception
of the NL$\sigma$Ms (\ref{Z}).
Accordingly, a perturbative analysis using the $\epsilon$ expansion from two dimensions was applied and 
suggested that
the $\beta$ functions for the dimensionless conductance $g=E_c/\Delta\equiv 1/2\pi t$ 
(or $1/\pi t$ for the symplectic case)
\cite{Weg89},
\beq
\beta(t)=\epsilon t +
\left\{
\begin{array}{ll}
-2t^2-12\zeta(3)t^5+\cdots & (\mbox{orthogonal})\\
-2t^3-6t^5+\cdots & (\mbox{unitary})\\
+t^2-\frac34\zeta(3) t^5+\cdots & (\mbox{symplectic})
\end{array}
\right. ,
\label{betafun}
\eeq
is likely to possess an IR-unstable fixed point that separates the metallic and insulating regimes
for the orthogonal and unitary classes in three or larger dimensions,
and the fixed point for the symplectic class persists even in two dimensions.
Once the presence of a fixed point is assumed,
the energy spectrum close to the thermodynamic limit
is expected to split clearly into the `metallic' region in the band center 
the `insulating' region at the band edges, depending on the mean level spacing.
The energy levels should obey the statistics of random matrices
and all eigenfunctions are extended in the former,
whereas in the latter all eigenfunctions are localized and accordingly the 
energy levels should have no correlation (Poisson statistics).
As the disorder strength is increased, 
the boundary of two regions, called {\it mobility edge},
move toward the band center and disappear alongside the extended eigenstates between two edges,
leading to the metal-insulator transition
\cite{KM93,MEGO}
Being an unstable boundary region separating Wigner-Dyson and Poisson statistics,
the mobility edge is expected to exhibit an intermediate level statistics
associated with fractal wave functions,
corresponding to the NL$\sigma$M precisely at the IR-unstable fixed point.
Although the width of the mobility edge (in the physical unit) shrinks under an increment of the lattice size, 
such `critical' statistics \cite{SSSLS}
should be stable and
depend only on the fixed point value of the conductance
(which in turn depends on the dimensionality $d=2+\epsilon$),
and possibly on the boundary condition and the aspect ratio of the lattice \cite{SP}.
It should otherwise be universal in a sense that
it originates from fine-tuning of
a single relevant coupling constant (conductance) and all other irrelevant couplings should play no role
\cite{KM81}.
These expectations have been verified numerically on the lattice
\cite{Eva,Hof96,KSOS}. 
For that purpose it is customary to use the uniform on-site randomness $\epsilon_{\rm r}$ of width $2W$
for a practical reason that the mean level density has a plateau in the band center,
which attains maximal efficiency in the
spectral averaging (as compared to Gaussian randomness that is theoretically easier to deal with).
In Fig.2 we plot the LSDs
from the central plateaux of the spectra of Anderson Hamiltonians 
at $W=16.4$ and $\alpha=0$ for the orthogonal case \cite{Eva} and
$W=18.1$ and $\alpha=0.2$ for the unitary case \cite{Hof96},
on cubic lattices of various sizes, with periodic boundary conditions on all sides.
\begin{figure}[h]
\centering
\includegraphics[scale=0.82,bb=0 0 260 170]{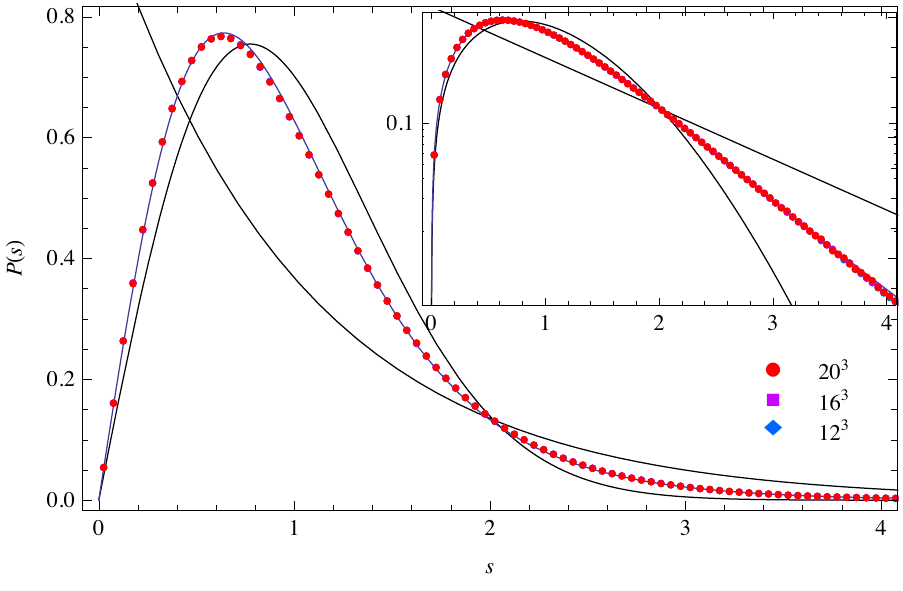}~~
\includegraphics[scale=0.82,bb=0 0 260 170]{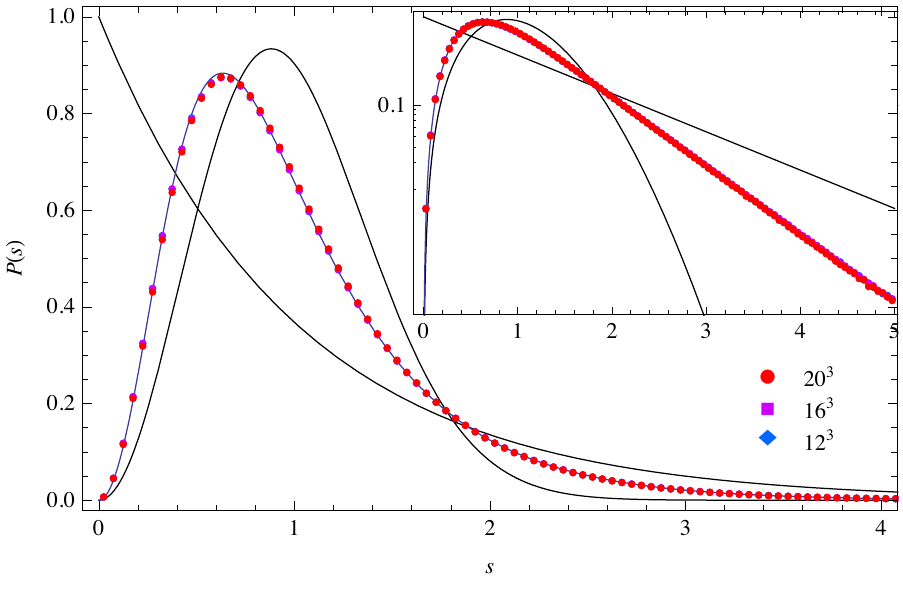}\\
\caption{
LSDs from the mobility edge of Anderson Hamiltonians of
orthogonal (left) and unitary (right) classes (red dots).
Parameters of Anderson Hamiltonians are: 
lattice size $L^3=12^3, 16^3, 20^3$, \#{}${\rm samples}=2\cdot10^5, 8\cdot10^4, 2\cdot10^4$,
eigenvalue window $\lambda\in[-2,2]$,
with uniform randomness of width $W=16.4$, (orthogonal), and
$W=18.1$, $\alpha=1.25$,  (unitary).
Best-fitting LSDs of deformed RMs at
$a=1.664(9)$ (orthogonal) and $a=3.572(13)$ (unitary)
are plotted in blue, and the LSDs of GOE, GUE, and Poisson in black.}
\end{figure}
We observe that the LSDs
from these eigenvalue windows
are indeed scale invariant, and are in between and a hybrid of 
Wigner-Dyson and Poisson,
\beq
P_{\beta}(s)\sim {\rm cst.}\,
s^\beta\ \ (s\ll 1),\ \ \ \log P_\beta(s)\sim -{\rm cst.}\, s\ \ (s\gg 1).
\label{Pscrit-asympt}
\eeq
This observation signifies that the whole band center
at these fine-tuned values of parameters
indeed belong to the mobility edge
corresponding to the presumed IR-unstable fixed point of the NL$\sigma$M.
We note that scale-invariant critical statistics is also observed for the
two-dimensional Anderson Hamiltonian in the symplectic class \cite{ASO},
but {\it not} in other two classes, in accordance with the
perturbatively $\beta$ functions (\ref{betafun}).

The relationship between the anomalous behavior of the energy eigenfunctions and
the fluctuation of the energy levels is widely accepted as follows: 
if the eigenfunctions from an energy window are multifractal \cite{SG}, i.e.,
the inverse participation ratio scales as
\beq
\sum_{\rm r} \left\langle |\psi_\lambda({\rm r})|^{2p} \right\rangle \propto L^{-D_p(p-1)}
\label{MF}
\eeq
with $D_p$ in between localized ($D_p=0$) and extended ($D_p=d$) states, 
then two such eigenfunctions overlap only sparsely,
$\sum_{\rm r} \left\langle |\psi_{\lambda}({\rm r})|^{2} |\psi_{\lambda'}({\rm r})|^{2}
\right\rangle \propto |\lambda-\lambda'|^{-(1-D_2)/d}$
for $|\lambda-\lambda'|\gg \Delta$.
Since only distant levels become less repulsive and less rigid,
it modifies the tail of the LSD to quasi-Poissonian,
whereas the small-$s$ behavior is not much affected, leading to (\ref{Pscrit-asympt}). 

\section{Deformed random matrices}
Once the existence of critical statistics is established,
an immediate challenge is to
derive  analytically
its statistical distributions,
such as the LSD and the two-level correlation function.
Since the NL$\sigma$M (\ref{Z}) that originates from the microscopic theory 
can be solvable only for very exceptional quasi-1D cases
for which Duistermaat-Heckman localization theorem is applicable \cite{LSZ},
one natural path
is to find a solvable effective model based on the symmetry and universality arguments.
One might wonder how the Anderson localization transition, for which the dimension
of the system is crucial, could possible be modeled by some effective RM ensemble,
which obviously carries no information on the dimensionality.
An answer to this frequently asked question is that it is the series of fractal dimensions
$D_p$ in (\ref{MF}) that dictates the level statistics (such as level repulsion and rigidity)
through the overlap of eigenfunctions,
and not the actual dimension $d$ of the system.
Accordingly, an effective model that possesses the crucial fractal property (\ref{MF})
and that reduces to the classical random matrices when the deformation is turned off,
might as well reproduce the critical statistics.
In this spirit, three seemingly different ensembles of random matrices have been proposed:
(I) Stieltjes-Wigert random matrices with 
a probability measure $dH\,\exp\{-(1/2a)(\log H)^2\}$ \cite{MCIN}
(termed as `nonclassical ensemble' in \cite{For}),
(II) power-law banded random matrices \cite{MFDQS}, and
(III) 1D free fermions at finite temperature \cite{MNS}.
Instead of going deeply into each model,
we merely mention the following key properties and refer the details to the original articles:
\begin{itemize}
\item
Ensemble I is a manifestly invariant ensemble.
Through an unusual unfolding $\lambda={\rm e}^{2a x}$
that results from the mean eigenvalue density $\bar{\rho}(\lambda)=1/2a\lambda$
and the cumulative density $x=\int d\lambda \bar{\rho}(\lambda)=(1/2a) \log\lambda$,
any $k$-level correlator for the unitary class is expressed as $\det [K_a(x_i,x_j)]_{i,j=1}^k$
in terms of a deformed spectral kernel
\beq
K_a(x,x')=\frac{\sin \pi(x-x')}{(\pi/a) \sinh a(x-x')},
\label{sinsinh}
\eeq
for a range of the deformation parameter $a$ for which the level
correlation is approximately translation-invariant.
\item
Ensemble II has a fixed preferred basis 
$U={\rm diag}\left\{\e^{2\pi i \ell/N}\right\}_{\ell=1}^{N}$ in a sense that the probability measure of
the matrix $H$ contains an extra factor $\exp(-b\,{\rm tr}\left|[H,U]\right|^2)$ that favors $H$ aligned to $U$
(in addition to the conventional Gaussian weight $\exp(-{\rm tr}\,H^2)$), and possesses the property (\ref{MF}).
Being Gaussian, it is directly mapped to a 1D NL$\sigma$M.
For the unitary class, the connected two-level correlator derived from the supersymmetric version
of the NL$\sigma$M (\ref{Z}) takes the form $-K_{1/2g}(x,x')^2$
for $g\gg1$, to the order $O(g^{-2})$ \cite{ASA}.
There
the $\sinh$ function originates from
the spectral determinant of the diffusion operator in (\ref{Z}),
\beq
\prod_{n\geq 1}\frac{1}{1+({s}/{2\pi g n})^2}=\frac{s/2g}{\sinh (s/2g)}.
\eeq
This connection relates the phenomenologically introduced parameter $a$ in
Ensemble I to a physical quantity (conductance) $g$ by $a=1/2g$.
\item
Ensemble III is a modified version of Ensemble B, obtained by replacing the constant $U$
for an integration over a group manifold (${\rm U}(N)$ in the original version,
recovering the invariance of GUE).
Being equivalent to a system of free fermions at a temperature related to $b$,
the probability distribution of particles' loci
(i.e.\ of eigenvalues of random matrices) is readily determined,
leading again to the same connected two-level correlator $-K_a(x,x')^2$.
\end{itemize}
Entrusting that these universality among three different ensembles
\cite{KM}
to be an indication of uniqueness of multifractal deformation of
the classical random matrices,
one of the authors (SMN) computed the LSDs of Ensemble I
in three symmetry classes \cite{Nis98,Nis99} (Fig.3).
Ensemble I is most suited for analytical computation of eigenvalue correlation,
because established techniques for invariant RM ensembles
(such as Tracy-Widom method of evaluating the Fredholm determinant
${\rm Det}(I-K_a\chi_{[0,s]})$ \cite{TW94}, and the relationships between $\beta=2$ and $\beta=1,4$
\cite{TW96})
are applicable, only with a replacement of the kernel from (\ref{sin}) to (\ref{sinsinh}).
For e.g.\ the unitary class, the LSD is expressed as
$P_{\beta=2}(s)=\partial_s^2 \e^{-\int_0^s dt R(t)}$ in terms of the diagonal resolvent 
$R(t)=\left\langle t|  K_a(I-K_a)^{-1} |t\right\rangle$, which satisfies
a transcendental equation of Painlev\'e VI type \cite{Nis98},
\bea
&&
R'(t)\left( [a R(t)]^2 +a\sinh 2at\, R(t)R'(t)
+[\sinh at\,R'(t)]^2 \right)
\nonumber\\
&&=
\left[ a \cosh at\,R'(t)  + \frac{\sinh at}{2}R''(t) \right]^2
+ [\pi \sinh at\,R'(t)]^2,
\label{PVI}
\eea
under the boundary condition $R(0)=R'(0)=1.$
\begin{figure}[t]
\centering
\includegraphics[scale=1.3,bb=0 0 166 110]{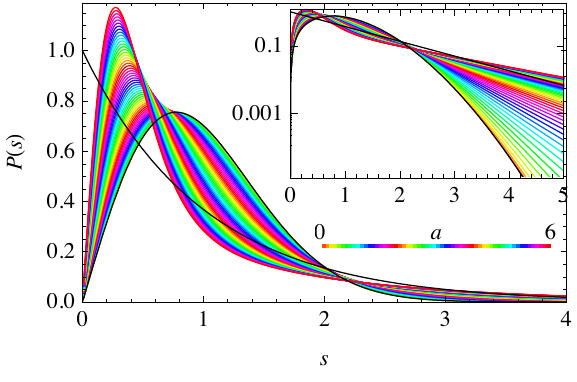}~
\includegraphics[scale=1.3,bb=0 0 166 114]{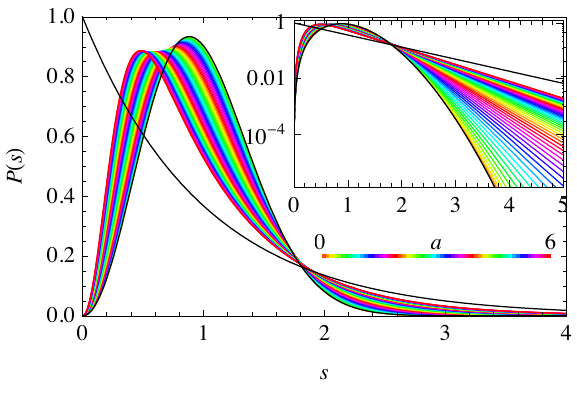}\\
\includegraphics[scale=1.3,bb=0 0 166 110]{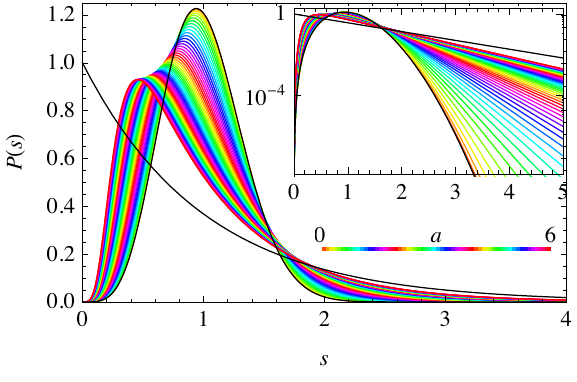}
\caption{
LSDs of
deformed RMs of orthogonal (above, left), unitary (above, right), and symplectic (below) classes.
The range of deformation parameters is $0.1 \leq a \leq 6.0$, taken at every 0.1.
These LSDs deviate from the black curves indicating the
LSDs of GOE, GUE, and GSE, respectively, toward Poisson distribution.
}
\end{figure}
Eq.~(\ref{PVI}) is a natural extension of the Painlev\'e V type equation \cite{JMMS}
derived for the resolvent of the sine kernel (\ref{sin}) at $a=0$.
One of the authors (SMN) then confirmed that the LSDs
at the mobility edge are well fitted, with a single tunable parameter $a$,
to these analytic formulas from the deformed kernel (blue curves in Fig.2)
for three symmetry classes \cite{Nis99}.
The $\chi^2/$dof of the fitting in the range $0\leq s \leq 5$ (with bin-size $0.05$) 
is as small as 0.23 (orthogonal) and 0.16 (unitary) for the case of $L^3=16^3$ and $n_{\rm conf}=8\cdot 10^4$.
This extremely precise matching
not merely justifies the validity of the deformed random matrices 
as an effective model of critical statistics a posteriori, 
but can even be used as a criterion for an eigenvalue window in the spectrum of a disordered spectra 
to belong to its mobility edge that separates extended and localized states.
Thus we move on to apply this criterion to the Dirac spectrum of QCD at the physical point,
in the high temperature phase.

\section{QCD Dirac spectra above T$_{\rm c}$}
The nature of Dirac eigenstates associated with small eigenvalues in the
chirally symmetric phase (in which $\langle \bar{q}q\rangle=-\pi
\bar{\rho}(0)/V=0$) has been debated since the appearance of Ref.\cite{HV},
partly motivated by the understanding of the fate of ${\rm U(1)}_A$ when
SU$(N_F)_A$ is restored.  The observation that low-lying eigenvalues at
$T\simeq T_c$ neither obey the Airy statistics of the `soft' band edge
of classical random matrices \cite{DHNR,FdHHLS}, nor the statistics of
random matrices at the (multi)critical edge \cite{JSV,ADMN}, left open the
issue of characterizing these eigenvalues and associated eigenstates.

An important step was undertaken by Garc\'ia-Garc\'ia and Osborn
who claimed that, right at the temperature of chiral symmetry restoration, the
LSD from the spectral window near the origin becomes stable under the
increment of spatial size 
and takes an intermediate form 
in
between Poisson and Wigner-Dyson statistics
\cite{GarciaGarcia:2006gr}. From this finding they speculated that Anderson
localization is the microscopic mechanism for the chiral phase
transition. However, they could not explicitly confirm a transition in the Dirac
spectrum from Poisson to Wigner-Dyson statistics. In retrospect, the reason
for that was partly the lack of enough statistics that made it necessary to
average spectral properties over spectral windows too wide. Another reason was
that around $T_c$ the density of localized modes is very low and one needs
large volumes to observe clear Poisson statistics.

In order to circumvent those issues,
subsequently some of 
the authors
performed simulations on much
larger lattices and at temperatures sufficiently above $T_c$
\cite{Kovacs:2009zj,Kovacs:2010wx,Kovacs:2012zq}.  
By choosing the spectral window
small enough for the spectral averaging to be justifiable, clear signs of the
mobility edge were observed
from the scale invariance of the LSD, 
even at small quark masses.  The location of the mobility edge
at a physical energy scale (well above the light quark mass) is seen to be
stable under the increment of the spatial lattice size, indicating that a
finite fraction of Dirac eigenstates per unit volume are localized in the
thermodynamic limit.

\begin{table}[h]
\begin{tabular}{lllllllll}
$N_C$ & $N_F$ & $\beta$ & $a$[fm] & $L_s$ & $L_t$ & $T$ & $n_{\rm conf}$ &
  $n_{\rm EV}$ \\ 
\hline 2 & 0 & 2.60 & - & 16, 24, 32, 48 & 4 & 2.6$T_c$ & 3k
  & 256 \\ 
3 & $2+1$ & 3.75 & .125 & 24, 28, $\ldots$, 48 & 4 & 394MeV &
  7k$\sim$40k & 256$\sim$1k
\end{tabular}
\caption{Simulation settings and parameters. Symanzik improved action and
  2-level stout-smeared staggered Dirac operator are employed for the SU(3)
  dynamical 
  case, whereas naive Wilson action and staggered Dirac operator are used for
  the SU(2) quenched case. 
    }
  \label{tab:simpars}
\end{table}

The present study is based on two sets of lattice simulations, one quenched
with the gauge group SU(2) and a dynamical SU(3) simulation with 2+1 flavors of
stout smeared staggered quarks at the physical point. The parameters of the
simulations are summarized in Table~\ref{tab:simpars}. More details of the
quark action, scale setting and quark masses for the dynamical simulation can
be found in Refs.~\cite{Aoki:2005vt,Borsanyi:2010cj}.
In the following we show that around the mobility edge in the spectrum, the
unfolded level spacing distribution is well described by a deformed random
matrix model. For determining the local spectral statistics throughout the
spectrum, the eigenvalue windows are chosen as small as possible while
attaining sufficient statistics.
We tried spectral averaging over
one (i.e.\ no spectral
averaging) to two level spacings for the smallest lattice ($16^3\times 4$ for
SU(2) quenched), and of six to twelve level spacings for the largest lattice
($48^3\times 4$ for SU(3) dynamical), around each designated point 
$\bar{\lambda}$ in the spectrum.
The spacings are unfolded by their mean value $\Delta(\bar{\lambda})$ within each
window, $s:=(\lambda_{i+1}-\lambda_i)/\Delta(\bar{\lambda})$.
We plot the probability density
of $s$ with a bin-size of 0.05 for the interval $0\leq s \leq 4$, and find a
best-fitting LSD of deformed random matrix ensembles (in symplectic class for
SU(2) quenched and in unitary class for SU(3) dynamical) for these 80 values by
varying the deformation parameter $a$.
Arbitrariness in the choice of
the bin-size could be avoided by fitting the cumulative LSD to the deformed
random matrices, but in order to achieve acute sensitivity to the
deformation parameter we employed the LSD itself for fitting.
In Fig.~4 we
exhibit sample plots of LSDs from Dirac eigenvalue windows centered at some
$\bar{\lambda}$s, and the best-fitting LSDs of deformed random matrices.

\begin{figure}[h]
\includegraphics[scale=0.82,bb=0 0 260 168]{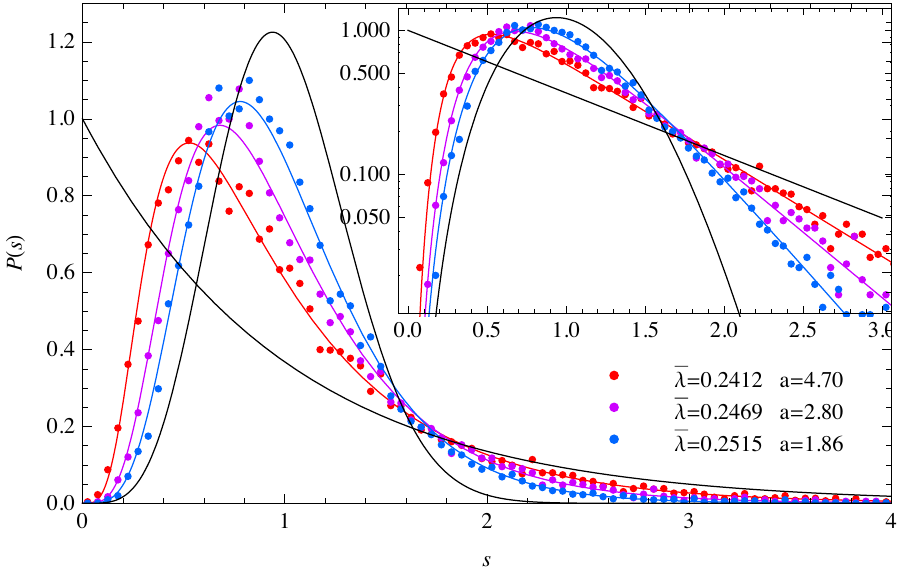}~~
\includegraphics[scale=0.82,bb=0 0 260 171]{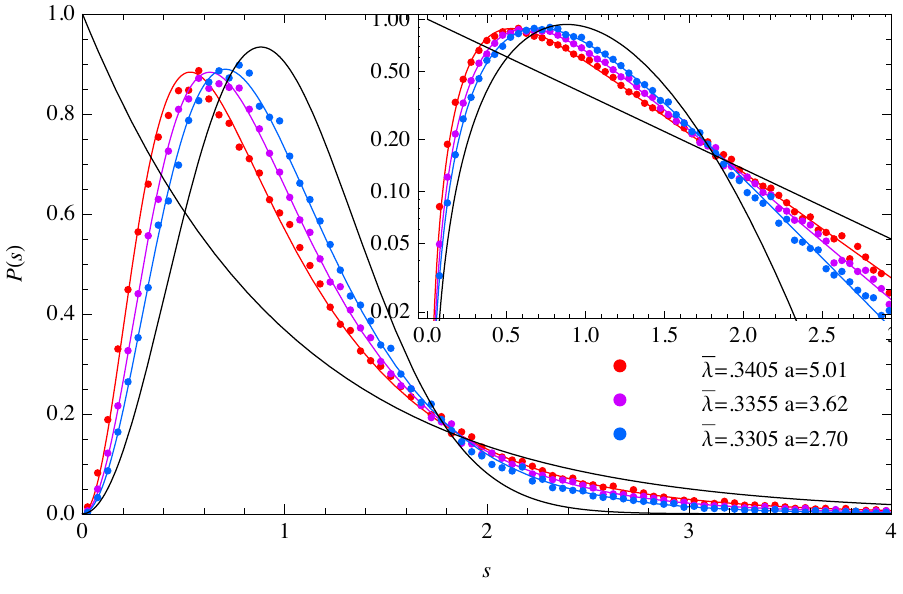}
\caption{LSDs from eigenvalue windows near the mobility edge, for SU(2)
  quenched
  (left) and for SU(3) dynamical
  (right).  LSDs of corresponding deformed RMs at the best-fitting 
  $a$ are plotted in the same color.}
\end{figure}

In the vicinity of the mobility edge (to be identified below), the results
from deformed random matrices fit quite nicely to the Dirac data, with
$\chi^2$/dof of around 1. As the LSD changes through the spectrum so does
the deformation parameter corresponding to the deformed random matrix ensemble
providing the best fit to the data.  In Fig.~5 we plot the best-fitting
deformation parameter $a$ versus the center of each eigenvalue window $\bar{\lambda}$.  
It turns out that the enveloping curves of $a(\bar{\lambda})$ for each lattice
size are rather insensitive to the number of the level spacings within the
window, so for the case of SU(3) dynamical
we show in Fig.5 left a plot with a
fixed number of LSs as: two for $L_s=24, 28$,
four for $L_s=32, 36, 40$, and six for $L_s=44, 48$.
\begin{figure}[h]
\includegraphics[scale=0.85,bb=0 0 260 162]{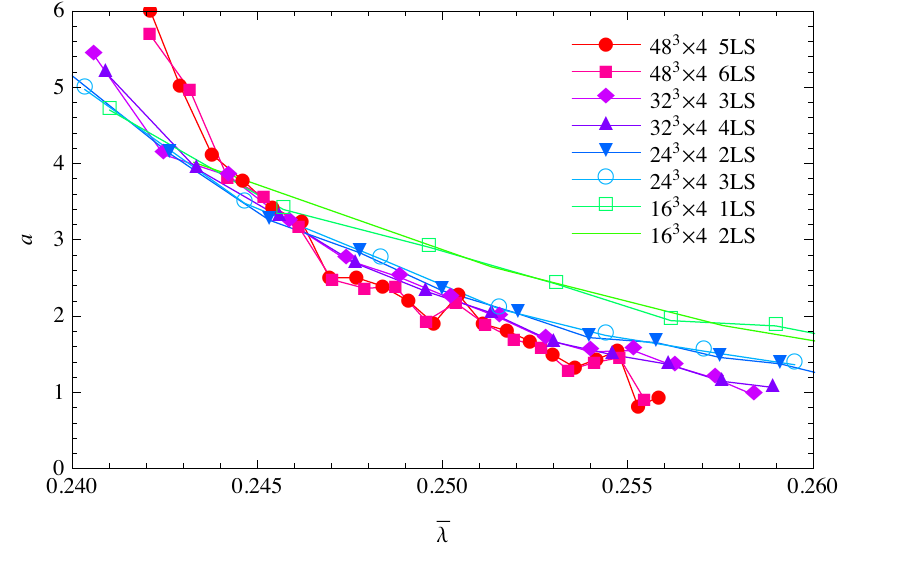}
\includegraphics[scale=0.85,bb=0 0 260 162]{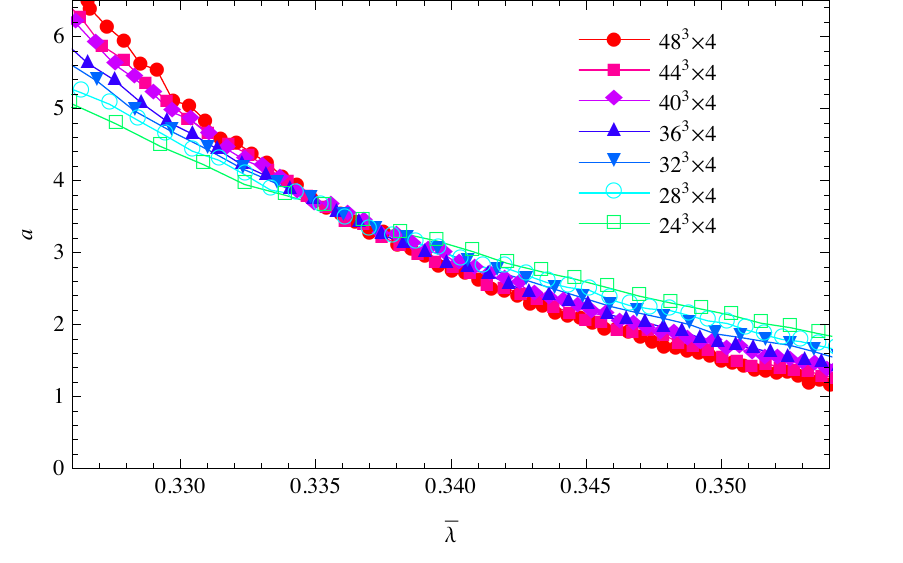}
\caption{
The best-fitting deformation parameter $a$ of deformed RMs versus
the center of the eigenvalue window $\bar{\lambda}$,
for SU(2) quenched
(left) and
for SU(3) dynamical
(right).
For each spatial lattice size, eigenvalue windows consisting of
various number of level spacings are used to determine the $a$ parameter.}
\end{figure}

From these figures one clearly sees that there are two regions in the
spectrum. For small $\lambda$, the deformation parameter, $a(\bar{\lambda})$,
increases with the system size and is expected to go to its Poisson limit as
$L_s\rightarrow \infty$. In this part of the spectrum, modes are
localized. For large $\lambda$ the deformation parameter decreases as the
system becomes bigger and $a(\bar{\lambda})$ is expected to go to its
Wigner-Dyson statistics limit as $L_s\rightarrow \infty$. This corresponds to
delocalized modes.  In between, there is a fixed point where
$a(\bar{\lambda})$ is independent of the lattice size. We identify this point
with the mobility edge separating localized and delocalized eigenmodes in the
thermodynamic limit.

By minimizing the variance of $a(\bar{\lambda})$ among various $L_s$, the
location of this mobility edge is determined as $\lambda_c a=0.245$ for SU(2)
quenched and $\lambda_c a=0.3353(3)$ for SU(3) dynamical.  The latter translates
to $\lambda_c=529$MeV in physical unit.  Furthermore, LSDs precisely from
tiny eigenvalue windows including $\lambda_c$ are scale invariant and
are all juxtaposed right on the top of the predictions from the deformed RMs
of symplectic class at $a=3.52$, and of unitary class at $a=3.64(6)$ (Fig.~6).
$\chi^2$/dof for the latter are $0.92\ (L_s=24) \sim  1.03 \ (L_s=48)$. 
We noticed, rather unexpectedly, that
the latter value of the deformation parameter $a$ at the mobility edge of the SU(3)
Dirac spectrum is consistent with the value $a=3.572(13)$ 
that corresponds to the mobility edge of the Anderson Hamiltonian in the unitary
class on an isotropic 3D lattice (Fig.~2, right). 

\begin{figure}[b]
\begin{center}
\includegraphics[bb=0 0 260 171]{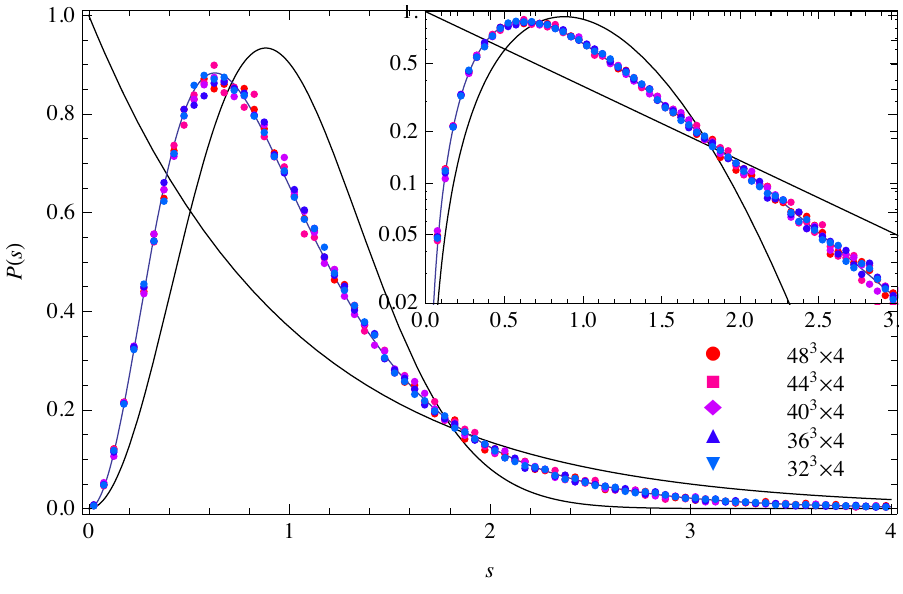}
\caption{LSDs from the eigenvalue window at the mobility edge $\lambda_c
  a=0.3355$ of Dirac spectrum for SU(3) dynamical, on $32^3\sim 48^3\times 4$
  lattices.  The best-fitting LSD of deformed unitary random matrix ensemble
  at $a=3.64$ is juxtaposed in a blue curve.}
\end{center}
\end{figure}

Finally we examine the shape parameters of the LSDs.  A customary choice in
the study of Anderson Hamiltonians is the variance of $s$ and the area up to
the crossing point ($s\simeq 0.5$) of Wigner and Poisson distributions,
$\upsilon=\int_0^\infty s^2 P(s) ds,\ \alpha=\int_0^{0.5} P(s) ds$.  In Fig.7
we plot the shape parameters $(\upsilon,\alpha)$ of LSDs from various
eigenvalue windows, on $(24^3\sim 48^3) \times 4$ lattices. 
One observes
that as the location of the window moves form the origin to the bulk, shape
parameters universally align on a specific curve connecting the
Poisson and Wigner-Dyson limits, 
regardless of the spatial size.  
Although this curve 
slightly
deviates from the one from the deformed unitary random matrices (red curve), 
these two cross at one point
(see the inset of Fig.7), precisely corresponding to the mobility edge.  This
findings may provide a further support on our claim that the mobility edge in
the Dirac spectrum in the high temperature phase survives the thermodynamic
limit, and its spectral fluctuation is described by the deformed random
matrices characterizing the fixed point of the NL$\sigma$M from
Anderson Hamiltonians.
\begin{figure}[h]
\begin{center}
\includegraphics[bb=0 0 260 166]{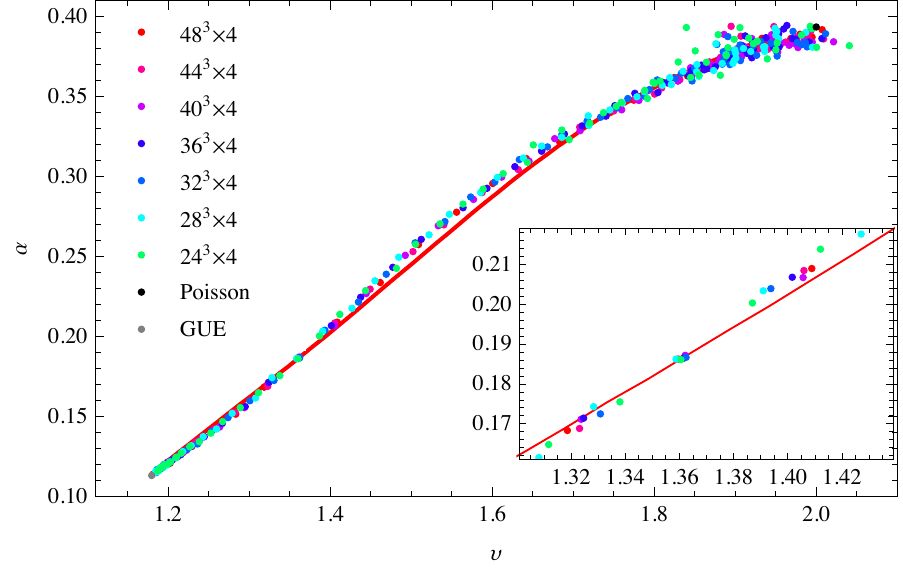}
\caption{Shape parameters of LSDs from various eigenvalue windows of SU(3) Dirac spectra,
and of the unitary deformed random matrices (red curve).}
\end{center}
\end{figure}

\section{Discussion: Critical statistics near the origin at T$_{\rm c}$}

We showed that the level spacing statistics across the Anderson-type
transition in the QCD staggered-Dirac spectrum can be described by the same deformed
RM ensemble that describes the transition in the Anderson
model. Together with the compatibility of the critical exponents of these two
models \cite{Giordano}, our result lends further support to the idea that
these two transitions indeed belong to the same universality class.
Further confirmation of our findings in the case of Ginsparg-Wilson
Dirac operators \cite{Kovacs:2009zj}, close to the physical point,
would be an important next step.

So far we discussed the transition in the spectrum at a fixed temperature
above $T_c$. To answer the original question of "whether chiral symmetry
restoration is driven by the Anderson localization of the quasi-zero modes of
the Dirac operator", one should approach the pseudo-critical temperature of
QCD from above, carefully monitoring the level statistics within each fine
spectral window.  By extrapolating the dependence of the mobility edge on the
temperature, two of us (TGK, FP) found that the mobility edge goes to zero
around $T\simeq 170$MeV \cite{Kovacs:2012zq}, which is compatible with the
pseudo-critical temperature of the QCD transition. 

It would be highly desirable to study how the spectral statistics changes as
the mobility edge goes to zero at $T_c$. In particular, it would be important
to verify that the scale invariant spectral statistics is still described by
the corresponding deformed random matrix model. There is, however, an
additional complication here. It is only in the bulk of the spectrum that
there is no difference between the chiral and the non-chiral version of the
given matrix model. However, as the temperature is lowered and $\lambda_c$
shifts to the spectrum edge, one has to use the chiral version of the matrix
model to describe the spectral statistics. 

For this purpose, Garc\'ia-Garc\'ia and Verbaarschot had already tailored a
chiral version of the deformed RM ensemble of type III \cite{GV}.  They have
obtained an exact form of the two-level correlation function $-K_a^{\rm
  chi}(x,x')^2$, and in the approximation of keeping $a^n (x-x')^n$ and
discarding $a^{n+2} (x-x')^n$ in the asymptotic limit $a\ll 1$ and $x, x'\gg
1$, it takes the form 
\beq K_a^{\rm chi}(x,x')\approx \frac{\pi}{2}\sqrt{x x'}
\left\{ \frac{J_1(\pi x) J_0(\pi x') - J_0(\pi x) J_1(\pi x')}{(1/a)\sinh
  a(x-x')} + (x'\to -x')\right\}.
\label{Kchi}
\eeq
This `deformed Bessel' kernel is the chiral counterpart of the 
`deformed sine' kernel $K_a(x,x')$ in
(\ref{sinsinh}) and reduces to this nonchiral deformed kernel in the limit $x,
x'\gg1$ with $x-x'$ fixed.  A practical problem of using the microscopic level
density $\rho(x)=K_a^{\rm chi}(x,x)$ for fitting the Dirac spectrum is that
$\rho(x)$ becomes rather structureless at a finite deformation parameter $a$
(see Fig.8 below) \cite{GT}.  On the other hand, the level number variance (an
integral transform of the two-level correlation function $-K_a^{\rm
  chi}(x,x')^2$) for large $x$ is sensitive to the parameter $a$, but one would
need extremely large lattices for the window at the origin containing dozens
of eigenvalues to be uniformly fitted with a single parameter $a$.

To circumvent this practical problem, we propose an alternative strategy of
using the $k^{\rm th}$ individual eigenvalue distributions
$p_k(x),\ k=1,2,\ldots$ \cite{DN01} that have characteristic peaky shapes of their own
and respond sensitively to the deformation, instead of the spectral density
that comprises of these peaks, $\rho(x)=\sum_{k\geq 1}p_k(x)$.  One could in
principle apply Tracy-Widom method to (\ref{Kchi}) to obtain a closed
analytic equation of Painlev\'e type for $p_k(x)$ analogous to (\ref{PVI}), but
for the actual fitting purpose it is sufficient to evaluate the Fredholm
determinant ${\rm Det}(I-K_a^{\rm chi}\chi_{[0,s]})$ and the resolvents ${\rm
  Tr}\left(K_a^{\rm chi}(I-K_a^{\rm chi})^{-1}\right)^k$ by the Nystr\"om-type approximation
\cite{Bor,Nis12}.  Here we employ the Gaussian quadrature of $100^{\rm th}$ order
and exhibit the
distributions of the five smallest eigenvalues as the deformation parameter
$a$ is varied in the range $0\leq a \leq 5$ (Fig.8).
\begin{figure}[h]
\centering
\includegraphics[bb=0 0 216 144]{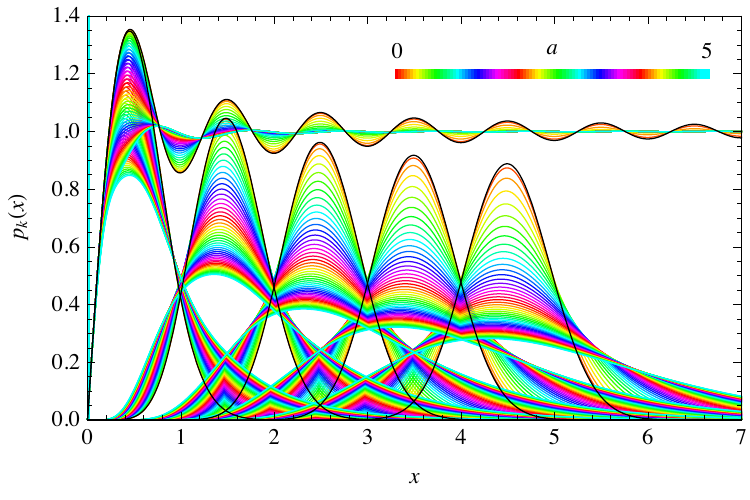}~
\includegraphics[bb=0 0 209 137]{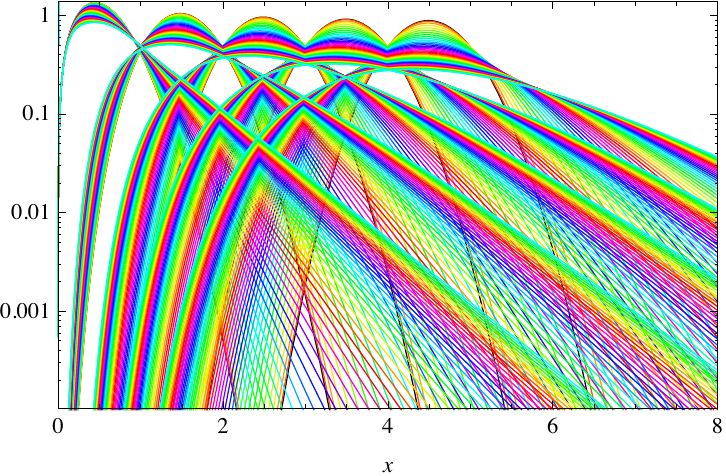}
\caption{
Linear (left) and logarithmic (right) plots of 
distributions of five smallest eigenvalues $p_1(x),\ldots,p_5(x)$ 
of the deformed chiral unitary RM ensembles.
The range of deformation parameters is $0.1 \leq a \leq 5.0$, taken at every 0.1.
These distributions deviate from $p_k(x)$ of chGUE (black curves)
toward Poisson distributions $(x^k/k!){\rm e}^{-x}$.
Corresponding microscopic level densities $\rho(x)$ (curves oscillating around 1) are 
also plotted in the left figure.
}
\end{figure}
Computational details, including applications to orthogonal and symplectic ensembles,
will appear elsewhere.
These distributions have Wigner-Dyson and Poisson asymptotics on different sides of the peaks,
\beq
\log p_k(x) \sim -{\rm cst.}\left(x-k+\frac12\right)^2\ \ (x< k-\frac12),\ \ 
\sim -{\rm cst.}\left(x-k+\frac12\right)\ \ (x> k-\frac12).
\eeq
Because of this very characteristic shapes,
they are unmistakable candidates for fitting the eigenvalues from the mobility edge occurring at the origin.
After verifying that eigenvalues in the mobility edge of the chiral Anderson Hamiltonian occurring
around the origin is described by these distributions,
and including the effect of quark masses in the $\varepsilon$ regime \cite{DN98,DN01},
we consider it a challenging but feasible task to compare the distributions of
each of smallest QCD Dirac eigenvalues at and around $T_{pc}$ and to answer to the point
quoted in the Introduction.\\

\noindent
{\bf Acknowledgements}:\ \ 
It is our great pleasure to thank the organizers of LATTICE$\;$2013 for the chance of presenting 
a work that was in progress and unpublished at the time of the conference, in the plenary session.
SMN thanks E. Itou and J.J.M. Verbaarschot for useful communications.

\end{document}